\documentclass[%
 reprint,
 amsmath,
 amssymb,
 aps,
 floatfix,
 pre,
onecolumn,
]{revtex4-2}
\usepackage{dcolumn}
\usepackage{amsmath,amssymb,amsfonts}
\usepackage{graphicx}
\usepackage{url}

\begin{document}

\title{Asymptotic Invariance of Kelly Allocation Under Power-Law Asset Dynamics: Evidence from Bitcoin}

\author{I.\ J.\ Vera Marun}
\affiliation{Department of Physics and Astronomy, University of Manchester, Manchester M13 9PL, United Kingdom} 

\date{\today}

\begin{abstract} 
	We examine log-optimal portfolio allocation when the long-run price of an asset follows a power-law trajectory, $P(t)=At^\alpha$, and its instantaneous return variance decays as $\sigma^2(t)=\sigma_0^2 t^{-2\gamma}$. Under a zero risk-free-rate benchmark, the continuous-time Kelly fraction scales as $K^\ast(t)=(\alpha/\sigma_0^2)t^{2\gamma-1}$. Exact temporal invariance therefore occurs when $\gamma=1/2$, whereas deviations from this value produce systematic age dependence in the allocation. We propose a scaling hypothesis connecting growth in network participation, effective market liquidity, and declining volatility. Under a specified set of scaling assumptions, this model predicts the benchmark exponent $\gamma=1/2$. Using historical daily Bitcoin prices, we estimate the power-law price exponent and examine the sensitivity of the volatility exponent to the length of the rolling window. For windows of four to nine years, the estimated volatility exponents have an arithmetic mean of 0.53 and a cross-window standard deviation of approximately 0.03. Because these estimates are obtained from overlapping observations and the same underlying price history, this spread is interpreted as a measure of model sensitivity rather than a formal confidence interval. Finally, we show that a time-dependent multiplicative contribution to return variance generally breaks exact Kelly invariance. We illustrate this result using a scenario in which Bitcoin transaction-fee variability affects the effective variance process. The results identify the conditions under which log-optimal allocation can remain stable under non-stationary power-law asset dynamics and clarify the assumptions required when applying this result to Bitcoin. 
\end{abstract}

\maketitle

\newpage

\section{Introduction}

Since the publication of Markowitz's mean--variance framework \cite{markowitz_portfolio_1952}, portfolio allocation models have frequently been applied using expected-return and variance parameters that are treated as constant over a specified investment horizon. Such an approximation may be unsuitable for assets whose long-run price and volatility exhibit systematic dependence on their age. This issue is particularly relevant in statistical-finance frameworks where volatility scaling and non-stationary market structure can significantly influence risk estimates over long horizons \cite{bouchaud_theory_2003}. Bitcoin provides a useful empirical case because previous studies have proposed a long-run power-law relationship between its price, network adoption, and absolute network age \cite{santostasi_mechanistic_2026}. Scaling behaviour has been observed in a wide range of economic and financial systems, motivating the search for universal descriptions of complex market dynamics \cite{stanley_scaling_2001}. More broadly, the econophysics literature has long emphasized scaling laws, heavy-tailed distributions, and emergent complexity in financial systems, suggesting that market dynamics may be fruitfully examined using concepts borrowed from statistical physics \cite{mantegna_introduction_1999}.

The occurrence of scaling laws across complex systems motivates the examination of Bitcoin as a non-stationary growing network rather than solely as a financial asset \cite{sornette_self-organized_2006}. Bitcoin prices also exhibit large cyclical departures from any smooth long-run trend. Log-periodic and discrete-scale-invariant models provide one framework for describing such departures \cite{sornette_discrete-scale_1998, wheatley_are_2019, feigenbaum_discrete_1996}. In this work, we do not estimate an LPPLS model or attempt to identify individual bubble regimes. Instead, we treat shorter-scale fluctuations, including those occurring near the approximately four-year supply cycle, as a source of sensitivity in the estimation of long-run price and volatility exponents. 

A distinction should be made between two different forms of scaling behaviour. Discrete-scale-invariant and log-periodic models describe oscillatory departures from a long-run trend and are commonly associated with speculative episodes, herding dynamics, and bubble-like behaviour \cite{sornette_discrete-scale_1998, wheatley_are_2019, feigenbaum_discrete_1996}. The present work focuses instead on the continuously scaled background component of the volatility process, represented by a smooth power-law envelope of the form $\sigma(t)\propto t^{-\gamma}$. In this interpretation, discrete-scale effects act as fluctuations around the underlying trend rather than determining the trend itself. The empirical objective is therefore to estimate the exponent of this continuous component and assess whether it is consistent with the benchmark prediction $\gamma=1/2$.

The purpose of this paper is to determine how log-optimal allocation depends on system age when both expected return and variance evolve according to power laws. Our contribution has three parts. First, we derive the general age dependence of the continuous-time Kelly fraction when the baseline asset price satisfies $P(t)=At^\alpha$ and return variance satisfies $\sigma^2(t)=\sigma_0^2t^{-2\gamma}$. This result shows that the allocation is independent of system age if and only if $\gamma=1/2$ under the assumed price dynamics. 

Second, we propose a scaling hypothesis connecting cumulative network growth, active market participation, effective liquidity depth, and volatility. The hypothesis provides one possible mechanism through which $\gamma=1/2$ could arise, but its intermediate scaling relationships are treated as explicit modelling assumptions rather than universal network laws. 

Third, we examine whether historical Bitcoin price data are empirically consistent with a volatility exponent near this benchmark. We quantify the sensitivity of the estimated exponent to the rolling-window length and distinguish this model sensitivity from formal sampling uncertainty. We then consider a general time-dependent variance factor, illustrated by a transaction-fee-dependent scenario, to show how exact allocation invariance can break down.

\section{A Scaling Hypothesis for Volatility Decay} 
\label{sec:scaling_hypothesis} 

We now present a scaling hypothesis that relates cumulative network growth to the evolution of effective market liquidity and volatility. The purpose of this section is not to claim a unique microstructural derivation of the volatility exponent. Rather, it identifies a set of assumptions under which the benchmark value $\gamma=1/2$ emerges. 
The long-run attenuation of Bitcoin volatility has been discussed in studies of Bitcoin valuation, adoption dynamics, and speculative market behaviour \cite{wheatley_are_2019,santostasi_mechanistic_2026}. More generally, volatility scaling, liquidity provision, and market-impact relationships have been central topics in the econophysics and statistical-finance literature \cite{mantegna_introduction_1999, bouchaud_theory_2003, bouchaud_trades_2018}. However, these studies generally treat volatility decay as an empirical regularity rather than as a quantity derived from explicit scaling relationships between network growth, participation, liquidity, and market impact. The objective of the present section is therefore not to establish that volatility declines, but to examine whether a simple scaling framework can produce the benchmark prediction $\gamma=1/2$ and thereby connect the observed decline to an underlying growth process.

Let the cumulative network measure scale with absolute system age as 
\begin{equation} 
	N_{\mathrm{wallets}}(t) \propto t^\eta , \label{eq:wallet_scaling_general} 
\end{equation} 
where $\eta>1$. Recent work on Bitcoin network adoption motivates the benchmark value $\eta\simeq3$ \cite{santostasi_mechanistic_2026}. We treat this value as an empirical input to the scaling model rather than as a universal exponent. 

We next assume that the flow of participants contributing to market activity per unit time scales with the growth rate of the cumulative network measure: 
\begin{equation} 
	N_{\mathrm{active}}(t) \propto \frac{dN_{\mathrm{wallets}}(t)}{dt} \propto t^{\eta-1}. \label{eq:active_scaling_general} 
\end{equation} 
Equation~(\ref{eq:active_scaling_general}) is a modelling assumption. Cumulative wallet counts, newly active participants, trading accounts, and order-book participants are not identical empirical quantities. The relationship should therefore be understood as a scaling approximation that can be tested using activity and market-microstructure data. 

To close the model, suppose that effective liquidity depth $L(t)$ grows sublinearly with active participation: 
\begin{equation} 
	L(t) \propto N_{\mathrm{active}}(t)^\delta \propto t^{\delta(\eta-1)}, 
	\label{eq:liquidity_scaling_general} 
\end{equation} 
where $\delta>0$ characterises how changes in active participation translate into market depth. Empirical studies of market microstructure and price impact motivate sublinear relationships between traded activity, liquidity, and price response \cite{lillo_master_2003,doyne_farmer_what_2004, bouchaud_trades_2018}. These studies do not, however, uniquely determine $\delta$ for Bitcoin over multi-year horizons. 

Finally, assume that the volatility scale decreases with the square root of effective liquidity depth: 
\begin{equation} 
	\sigma(t) \propto L(t)^{-1/2}. 
	\label{eq:vol_liquidity_relation} 
\end{equation} 
Combining Eqs.~(\ref{eq:wallet_scaling_general}) through (\ref{eq:vol_liquidity_relation}) gives
\begin{equation} 
	\sigma(t) \propto t^{-\delta(\eta-1)/2}. 
	\label{eq:volatility_general_scaling} 
\end{equation} 
Writing $\sigma(t)=\sigma_0t^{-\gamma}$ therefore yields 
\begin{equation} 
	\gamma=\frac{\delta(\eta-1)}{2}. 
	\label{eq:gamma_general} 
\end{equation} 

For the benchmark choices $\eta=3$ and $\delta=1/2$, the model predicts 
\begin{equation} 
	N_{\mathrm{active}}(t)\propto t^2, \qquad L(t)\propto t, \qquad \sigma(t)\propto t^{-1/2}, 
	\label{eq:benchmark_scaling} 
\end{equation} 
and hence 
\begin{equation} 
	\gamma=\frac{1}{2}. 
	\label{eq:unified_scaling} 
\end{equation} 

The value $\gamma=1/2$ is therefore a benchmark prediction conditional on the assumed exponents $\eta=3$ and $\delta=1/2$. Alternative relationships between cumulative adoption, active participation, liquidity depth, and volatility would generally produce a different value of $\gamma$. The decline in volatility may be viewed heuristically as an averaging effect associated with increasing effective market depth. We use this analogy only as motivation; the quantitative prediction follows from the explicit scaling assumptions above.

\section{Kelly Allocation Under Power-Law Drift and Variance} 
\label{sec:kelly_invariance} 

Consider a single risky asset with price $S(t)$ described locally by 
\begin{equation} 
	\frac{dS(t)}{S(t)} = \mu(t)\,dt+\sigma(t)\,dW(t), 
	\label{eq:sde_diffusion} 
\end{equation} 
where $\mu(t)$ denotes the expected instantaneous arithmetic return, $\sigma(t)$ is the instantaneous return volatility, and $W(t)$ is a standard Wiener process. We treat the deterministic long-run price trajectory 
\begin{equation} 
	P(t)=At^\alpha 
	\label{eq:price_attractor} 
\end{equation} 
as a model for the baseline expected price dynamics. Its proportional growth rate is 
\begin{equation} 
	\frac{1}{P(t)}\frac{dP(t)}{dt} = \frac{\alpha}{t}. 
	\label{eq:drift} 
\end{equation} 
Accordingly, we specify the baseline arithmetic return drift as 
\begin{equation} 
	\mu(t)=\frac{\alpha}{t}. 
	\label{eq:arithmetic_drift_assumption} 
\end{equation} 
This is a modelling specification for the expected return in Eq.~(\ref{eq:sde_diffusion}). It should be distinguished from the drift of $\ln S(t)$, which contains the usual It\^{o} correction $-\sigma^2(t)/2$. 

Let the instantaneous return variance satisfy the general power law 
\begin{equation} 
	\sigma^2(t)=\sigma_0^2 t^{-2\gamma}, 
	\label{eq:variance_general} 
\end{equation} where $\sigma_0^2>0$ sets the variance scale at $t=1$ and $\gamma$ is the volatility-decay exponent. 

For an unconstrained log-utility investor holding one risky asset and one risk-free asset, the continuous-time optimal risky fraction is 
\begin{equation} 
	K^\ast(t) = \frac{\mu(t)-r(t)}{\sigma^2(t)}, 
	\label{eq:kelly_operator_general} 
\end{equation} where $r(t)$ is the instantaneous risk-free rate \cite{merton_lifetime_1969,maclean_kelly_2011}. To isolate the age dependence associated with the risky asset, we use the benchmark $r(t)=0$. 

\textbf{Proposition 1 (Age dependence of log-optimal allocation).} 
\textit{Suppose that the expected instantaneous excess return and variance of an asset satisfy} 
\begin{equation} 
	\mu(t)-r(t)=a t^{-p}, \qquad \sigma^2(t)=b t^{-q}, 
	\label{eq:general_return_variance_powers} 
\end{equation} 
\textit{where $a,b>0$. Then the unconstrained log-optimal risky fraction is} 
\begin{equation} 
	K^\ast(t)=\frac{a}{b}t^{q-p}. 
	\label{eq:general_kelly_scaling} 
\end{equation} 
\textit{The allocation is independent of system age if and only if $p=q$.} 

\textit{Proof.} 
Substitution of Eq.~(\ref{eq:general_return_variance_powers}) into Eq.~(\ref{eq:kelly_operator_general}) gives 
\begin{equation} 
	K^\ast(t) = \frac{a t^{-p}}{b t^{-q}} = \frac{a}{b}t^{q-p}. 
\end{equation} 
The result is independent of $t$ if and only if the exponent satisfies $q-p=0$, or equivalently $p=q$. 
\hfill$\square$ 

For the power-law price and volatility model of Eqs.~(\ref{eq:arithmetic_drift_assumption}) and (\ref{eq:variance_general}), $a=\alpha$, $p=1$, $b=\sigma_0^2$, and $q=2\gamma$. The optimal fraction is therefore 
\begin{equation} 
	K^\ast(t) = \frac{\alpha}{\sigma_0^2}t^{2\gamma-1}. 
	\label{eq:kelly_general_gamma} 
\end{equation} 
Equation~(\ref{eq:kelly_general_gamma}) gives three regimes: 
\begin{align} 
	\gamma < \frac{1}{2} &\quad\Longrightarrow\quad K^\ast(t) \text{ decreases with system age}, \\
	\gamma = \frac{1}{2} &\quad\Longrightarrow\quad K^\ast(t)=\frac{\alpha}{\sigma_0^2}, \\
	\gamma > \frac{1}{2} &\quad\Longrightarrow\quad K^\ast(t) \text{ increases with system age}. 
\end{align} 

Thus, $\gamma=1/2$ is the condition for exact temporal invariance under the assumed drift model. The invariant allocation is a consequence of matching temporal exponents in expected excess return and variance. It does not require that the price exponent take any particular numerical value. 

The result applies to an idealised, unconstrained log-utility investor. It does not include leverage or position constraints, transaction costs, parameter uncertainty, jumps, drawdown limits, or estimation risk. Numerical values of $K^\ast$ should therefore be interpreted as model-implied benchmark allocations rather than investment recommendations.

\section{Empirical Estimates and Window Sensitivity} 
\label{sec:empirical} 

We examine whether historical Bitcoin price data are empirically consistent with the power-law dynamics introduced above. This exercise is an in-sample calibration and consistency check, rather than an out-of-sample test. 

Daily Bitcoin closing prices were obtained from CoinGecko \cite{noauthor_bitcoin_2026b}. The observations were ordered chronologically and system age was measured relative to the Bitcoin genesis date, January 3, 2009. Observations within the first year after genesis were excluded from the log-log regressions. Calendar dates missing from the source series were inserted and assigned the most recently available closing price before daily log returns were calculated. Because this procedure introduces zero returns on filled dates, the number of affected observations and the sensitivity of the results to omitting them should be reported as a robustness check. 

The full-sample price relation was estimated using 
\begin{equation} 
	\ln P(t)=\ln A+\alpha\ln t+\varepsilon_P(t). 
	\label{eq:price_regression} 
\end{equation} 
This regression gives the descriptive scaling estimate $\alpha=5.185\pm0.023$, with $R^2=0.9126$. The quoted uncertainty is the native OLS standard error and does not account for residual serial correlation, structural changes, or uncertainty in the power-law specification. More generally, the identification and estimation of power-law behaviour in empirical data require careful consideration of model selection, finite sample effects, and alternative heavy-tailed distributions \cite{clauset_power-law_2009}. The present regression should therefore be interpreted as a descriptive scaling estimate rather than a complete statistical test of power-law behaviour.

Realised annualised volatility was calculated over rolling windows of length $T$, with $T$ varied from one to twelve years. For each window length, the variance relation was estimated as 
\begin{equation} 
	\ln \widehat{\sigma}_T^2(t) = \ln \sigma_{0,T}^2 - 2\gamma_T\ln t + \varepsilon_{\sigma,T}(t). 
	\label{eq:variance_regression} 
\end{equation} 
The rolling estimates overlap strongly in time. Consequently, adjacent values of $\widehat{\sigma}_T^2(t)$ are not statistically independent, and native OLS standard errors understate the full uncertainty. We report them as descriptive regression errors only. The variation of $\gamma_T$ across window lengths is treated as a window-sensitivity diagnostic rather than as a collection of independent estimates.

\begin{figure}[htbp]
	\centering
	\includegraphics[width=\columnwidth]{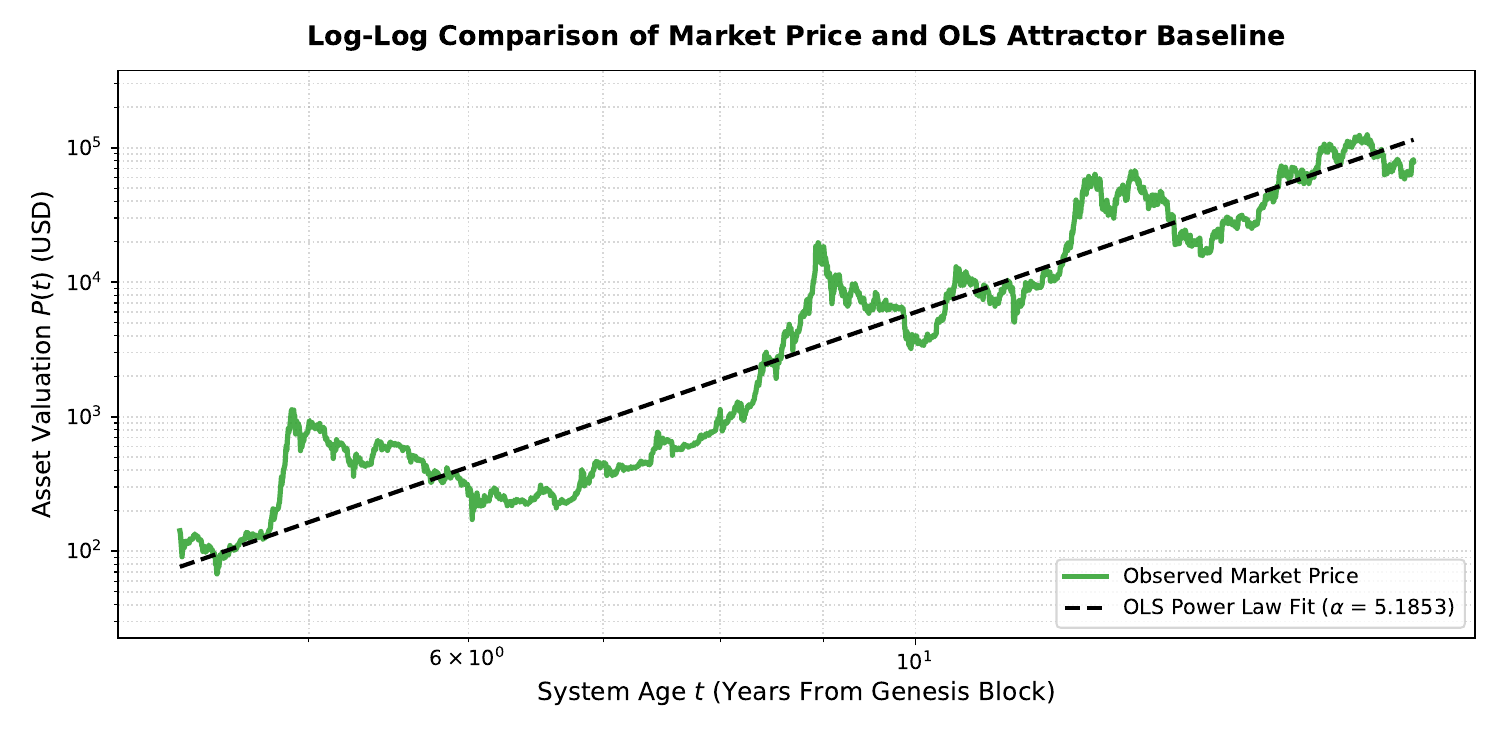}
	\caption{Historical daily Bitcoin closing prices and the fitted full-sample power-law trend, shown on logarithmic axes. The green curve represents the observed closing price as a function of system age, measured from the genesis date, and the black dashed line represents the OLS fit $P(t)=At^\alpha$, with $\alpha=5.1853$ and $R^2=0.9126$. Deviations from the fitted curve may contain cyclical, regime-dependent, and other non-power-law components; those components are not separately identified by this regression.}	
	\label{fig:price_regression}
\end{figure}

\begin{table}[htbp]
	\begin{center}
	\caption{Log-log OLS estimates of the realised-variance scaling relation $\ln \widehat{\sigma}_T^2(t) =\ln \sigma_{0,T}^2-2\gamma_T\ln t+\varepsilon_{\sigma,T}(t)$ for rolling-window lengths $T=1,\ldots,12$ years. The reported uncertainties are native OLS standard errors. Because the realised-variance observations are generated by strongly overlapping rolling windows, these errors do not account for serial dependence and should be interpreted as descriptive regression errors rather than formal confidence intervals. The final column gives the remaining span of system age available after the initial rolling window has been formed.}
	\label{tab:OLS} 
	\begin{tabular}{|c|c|c|c|c|}
	\hline 
	$T$ &  $\gamma_{\text{fitted}}$ & $\ln \sigma_0^2$ & $R^2$ & OLS range \\ 
	\hline 
	1 & $0.4576 \pm 0.0092\quad $ & $1.3310 \pm 0.0446$ & 0.3518 & 12.35 \\ 
	\hline 
	2 & $0.4382 \pm 0.0090\quad $ & $1.3278 \pm 0.0443$ & 0.3632 & 11.35 \\ 
	\hline 
	3 & $0.4765 \pm 0.0088\quad $ & $1.6034 \pm 0.0441$ & 0.4372 & 10.35 \\ 
	\hline 
	4 & $0.5240 \pm 0.0077\quad $ & $1.9414 \pm 0.0392$ & 0.5766 & 9.36 \\ 
	\hline 
	5 & $0.5316 \pm 0.0070\quad $ & $2.0598 \pm 0.0362$ & 0.6548 & 8.36 \\ 
	\hline 
	6 & $0.5260 \pm 0.0067\quad $ & $2.0864 \pm 0.0355$ & 0.6936 & 7.36 \\ 
	\hline 
	7 & $0.5459 \pm 0.0061\quad $ & $2.2408 \pm 0.0327$ & 0.7733 & 6.36 \\ 
	\hline 
	8 & $0.5577 \pm 0.0060\quad $ & $2.3454 \pm 0.0324$ & 0.8156 & 5.36 \\ 
	\hline 
	9 & $0.4786 \pm 0.0051\quad $ & $1.9446 \pm 0.0282$ & 0.8448 & 4.36 \\ 
	\hline 
	10 & $0.6188 \pm 0.0071\quad $ & $2.7191 \pm 0.0395$ & 0.8600 & 3.36 \\ 
	\hline 
	11 & $0.9151 \pm 0.0094\quad $ & $4.3972 \pm 0.0528$ & 0.9165 & 2.36 \\ 
	\hline 
	12 & $1.4037 \pm 0.0162\quad $ & $7.2197 \pm 0.0920$ & 0.9377 & 1.36 \\ 
	\hline 
	\end{tabular} 
	\end{center}
\end{table}

The estimated exponent depends materially on the rolling-window length. For $T=1$ year, the fitted value is $\gamma_T=0.458$, while the four-year window gives $\gamma_T=0.524$. The estimates for $T=4$ to $T=9$ remain in the approximate range $0.48$ to $0.56$. For longer windows, the remaining regression span becomes short relative to the total history, and the fitted exponent rises sharply. This behaviour is qualitatively reminiscent of finite-size effects that arise when the available observation horizon becomes comparable to the characteristic scale of the fitted process \cite{privman_finite_1990}.

The increasing $R^2$ and decreasing native OLS standard errors observed over part of this sequence should not be interpreted as unambiguous evidence of increasing statistical precision. Longer and more strongly overlapping windows generate smoother derived variance series, which can mechanically reduce regression residuals. At the same time, the number of effectively independent volatility observations and the remaining temporal span both decrease. 

We therefore use the variation with $T$ as a sensitivity analysis. The range $T=4$ to $T=9$ is reported because it includes the approximate four-year protocol cycle while retaining a regression span of more than four years. This criterion is diagnostic and should not be interpreted as identifying a unique optimal window.

\begin{figure}[htbp]
	\centering
	\includegraphics[width=\columnwidth]{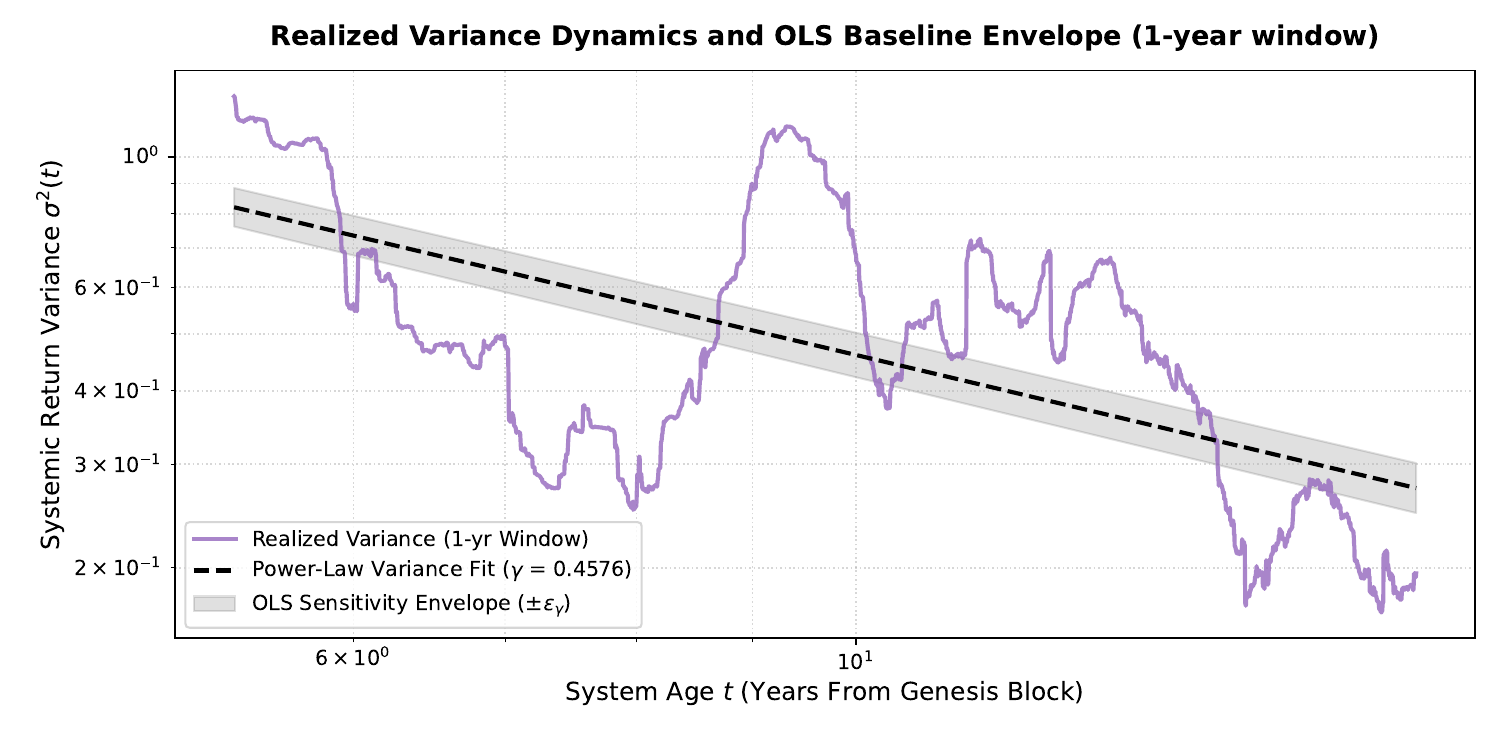}
	\includegraphics[width=\columnwidth]{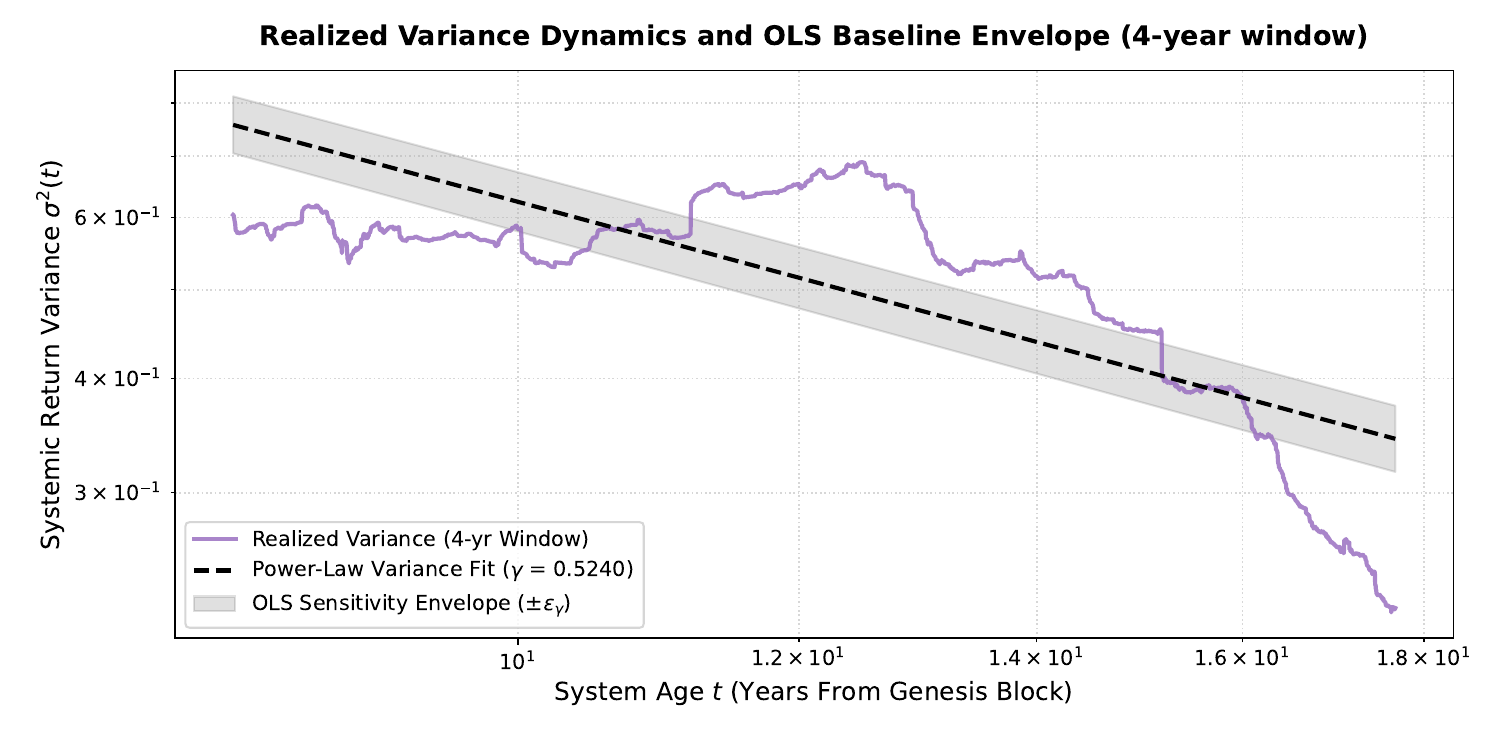}
	\caption{Realised annualised return variance and fitted power-law variance curves for rolling windows of $T=1$ year (top) and $T=4$ years (bottom). The purple curves show the variance estimates derived from overlapping rolling windows, and the black dashed lines show the corresponding OLS fits. The shaded regions are obtained by shifting the fitted slope and intercept by their native OLS standard errors. They are displayed as descriptive parameter-sensitivity envelopes and should not be interpreted as formal confidence or prediction bands because slope--intercept covariance and serial dependence are not included.}
	\label{fig:variance_decay}
\end{figure}

To summarise sensitivity to temporal coarse-graining, we consider the estimates obtained for $T=4$ to $T=9$ years. These estimates are derived from the same underlying price history and from overlapping transformations of that history. They are therefore dependent and are not treated as an independent statistical ensemble. 

Across the diagnostic window range T=4--9 years, the fitted volatility exponent has an arithmetic mean of 0.53 and a sample standard deviation of approximately 0.03. An inverse-variance weighted average using the native OLS regression errors yields a similar value ($\approx 0.52$); however, because the estimates are derived from overlapping windows and are interpreted here as a window-sensitivity analysis rather than an independent statistical ensemble, we report the arithmetic mean. The standard deviation measures sensitivity to the selected rolling window and is not a confidence interval for a population parameter. Within this diagnostic range, the estimates are broadly consistent with the benchmark value $\gamma=1/2$ predicted by the scaling hypothesis. 

The economically relevant implication of deviations from $\gamma=1/2$ is given directly by Eq.~(\ref{eq:kelly_general_gamma}). Between two system ages $t_1$ and $t_2$, the model-implied allocation changes by 
\begin{equation} 
	\frac{K^\ast(t_2)}{K^\ast(t_1)} = \left(\frac{t_2}{t_1}\right)^{2\gamma-1}. 
	\label{eq:kelly_horizon_ratio} 
\end{equation} 
Equation~(\ref{eq:kelly_horizon_ratio}) provides a more direct measure of practical allocation drift than the limiting behaviour of $dK^\ast/dt$. Exact invariance requires $\gamma=1/2$; estimates close to this value imply slow, but not necessarily zero, variation with system age.

\begin{figure}[htbp]
	\centering
	\includegraphics[width=0.92\columnwidth]{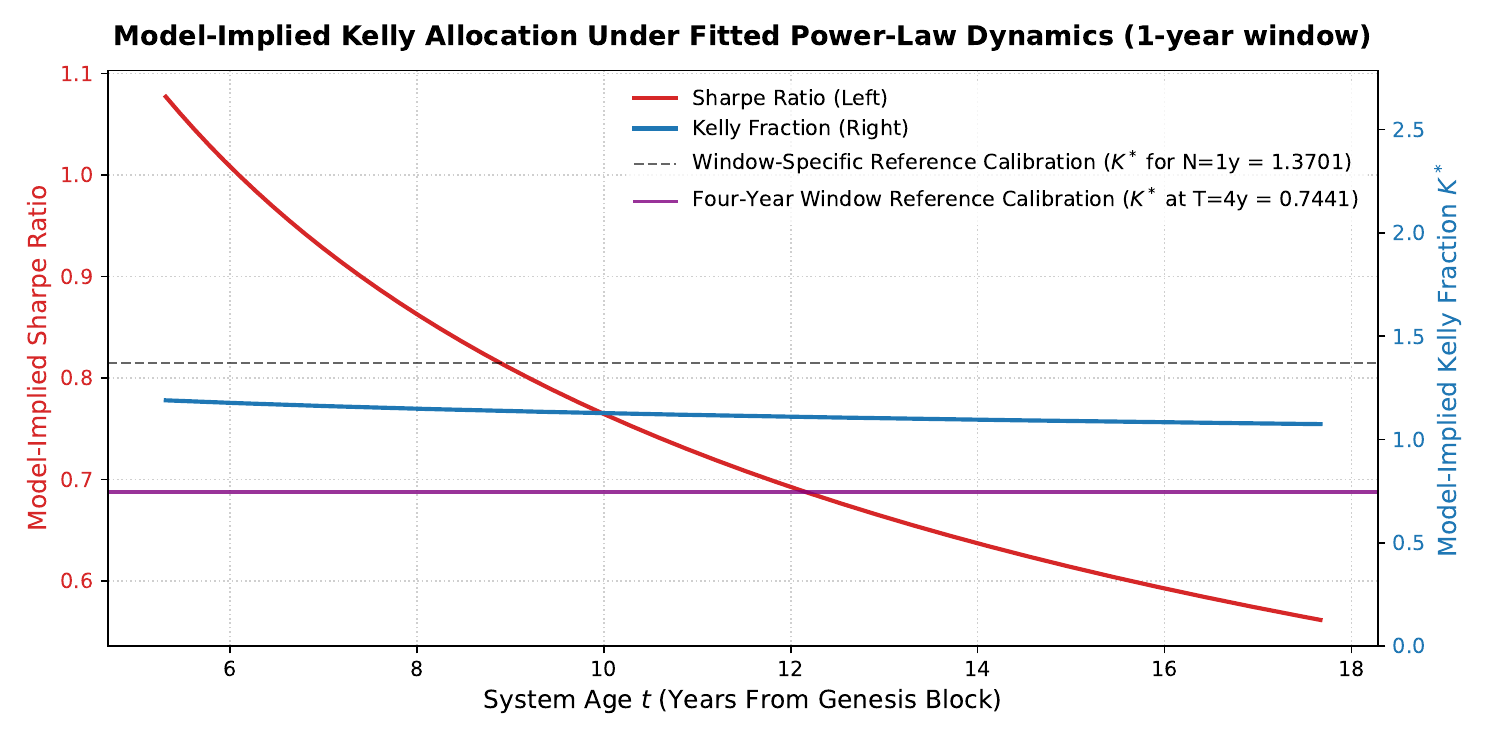}
	\includegraphics[width=0.92\columnwidth]{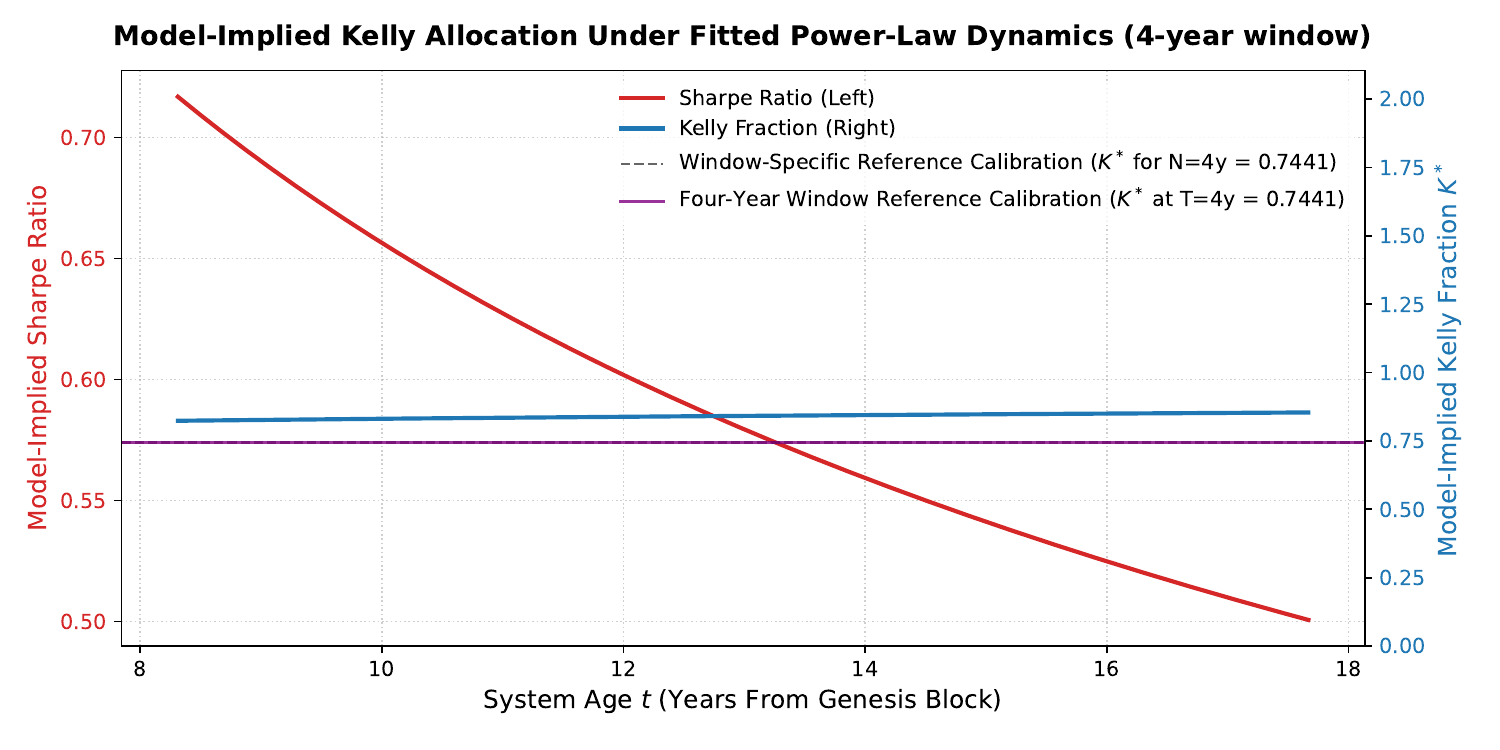}
	\caption{Model-implied Sharpe ratio and unconstrained Kelly fraction obtained from the fitted power-law parameters for rolling windows of $T=1$ year (top) and $T=4$ years (bottom). The curves are consequences of the fitted model and do not constitute an independent empirical test of allocation performance. The horizontal lines show calibration-specific reference values of $\alpha/\sigma_{0,T}^2$. Under the general model, the Kelly fraction scales as $K^\ast(t)\propto t^{2\gamma_T-1}$ and is exactly constant only when $\gamma_T=1/2$.}
	\label{fig:portfolio_invariance}
\end{figure}

The estimated variance intercept changes systematically with the choice of rolling-window length. This behaviour reflects the effect of temporal smoothing on the derived realised-volatility series. Larger windows suppress shorter-term fluctuations and therefore alter both the fitted exponent and intercept. Consequently, the estimated intercept should be interpreted as a calibration-dependent parameter rather than a unique structural quantity.

Using the four-year rolling-window calibration as an illustrative benchmark gives 
\begin{equation} 
	K^\ast_{T=4} = \frac{\alpha}{\sigma_{0,T=4}^2} = \frac{5.1853}{\exp(1.9414)} \approx 0.74. 
	\label{eq:kelly_four_year_calibration} 
\end{equation} 
This value is the model-implied unconstrained Kelly fraction under the four-year-window calibration. It is sensitive to the selected window, annualisation convention, data treatment, price-history range, and estimated variance intercept. It should not be interpreted as a uniquely unbiased allocation or as an investment recommendation. The principal result of the analysis concerns the temporal scaling of $K^\ast(t)$, rather than the particular numerical level obtained from one calibration.

\section{An Illustrative Breakdown of Invariance Under Time-Dependent Variance} 
\label{sec:breakdown} 

The invariance condition derived in Section~\ref{sec:kelly_invariance} requires the expected excess return and variance to have matching temporal exponents. It can also fail if variance contains an additional time-dependent factor that is not shared by the expected return. 

Consider the general variance process 
\begin{equation} 
	\sigma_{\mathrm{realized}}^2(t) = \sigma_0^2 t^{-1} g(t), 
	\label{eq:general_variance_modifier} 
\end{equation} 
where $g(t)>0$ represents an additional source of time-dependent risk. With $\mu(t)=\alpha/t$ and $r(t)=0$, the corresponding Kelly fraction is 
\begin{equation} 
	K^\ast_{\mathrm{realized}}(t) = \frac{\alpha}{\sigma_0^2}g(t)^{-1}. 
	\label{eq:kelly_general_modifier} 
\end{equation} 
Exact temporal invariance is recovered only when $g(t)$ is constant. 

As an illustrative Bitcoin-specific scenario, suppose that variability in transaction-fee revenue affects the effective variance process. Define the relative fee dispersion 
\begin{equation} 
	f_{\mathrm{fee}}(t) = \frac{\sigma_{\mathrm{fee}}^2(t)} {\langle F_{\mathrm{fee}}(t)\rangle^2}, 
	\label{eq:fee_dispersion} 
\end{equation} 
and postulate the variance modifier 
\begin{equation} 
	g(t)=1+\xi f_{\mathrm{fee}}(t), \qquad \xi>0, 
	\label{eq:fee_variance_modifier} 
\end{equation} 
where $\xi$ is a response parameter. This relationship is a scenario assumption and is not estimated from the historical price data used in Section~\ref{sec:empirical}. 

The resulting variance and allocation are 
\begin{equation} 
	\sigma_{\mathrm{realized}}^2(t) = \frac{\sigma_0^2}{t} \left[ 1+\xi \frac{\sigma_{\mathrm{fee}}^2(t)} {\langle F_{\mathrm{fee}}(t)\rangle^2} \right], 
	\label{eq:subordinated_variance} 
\end{equation} 
and 
\begin{equation} 
	K^\ast_{\mathrm{realized}}(t) = \frac{\alpha}{\sigma_0^2} \left[ 1+\xi \frac{\sigma_{\mathrm{fee}}^2(t)} {\langle F_{\mathrm{fee}}(t)\rangle^2} \right]^{-1}.
	\label{eq:k_realized_fracture} 
\end{equation} 

Differentiating Eq.~(\ref{eq:k_realized_fracture}) gives 
\begin{equation} 
	\frac{dK^\ast_{\mathrm{realized}}(t)}{dt} = -\frac{\alpha}{\sigma_0^2} \xi \left[1+\xi f_{\mathrm{fee}}(t)\right]^{-2} \frac{df_{\mathrm{fee}}(t)}{dt}. 
	\label{eq:dk_dt_fracture} 
\end{equation} 
Thus, the allocation is time independent only if relative fee dispersion is constant. Any systematic time variation in $f_{\mathrm{fee}}(t)$ introduces corresponding variation in the model-implied Kelly fraction. 

This scenario illustrates a more general point: even when baseline variance decays as $1/t$, an additional time-dependent risk factor can prevent exact allocation invariance. Establishing whether transaction-fee dynamics produce such an effect would require a separate empirical model linking fee revenue, network security, market liquidity, and return variance.

\section{Conclusion} 

For an asset whose expected instantaneous excess return and variance follow power laws in system age, the continuous-time log-optimal allocation satisfies 
\begin{equation} 
	K^\ast(t)\propto t^{q-p}, 
\end{equation} 
where $p$ and $q$ are the decay exponents of expected excess return and variance, respectively. For a power-law price baseline $P(t)=At^\alpha$ and variance $\sigma^2(t)=\sigma_0^2t^{-2\gamma}$, this becomes 
\begin{equation} 
	K^\ast(t) = \frac{\alpha}{\sigma_0^2}t^{2\gamma-1}. 
\end{equation} 
Exact temporal invariance therefore occurs when $\gamma=1/2$. 

We proposed a scaling hypothesis connecting cumulative network growth, active participation, effective liquidity depth, and declining volatility. Under the benchmark assumptions $N_{\mathrm{wallets}}\propto t^3$ and $L\propto N_{\mathrm{active}}^{1/2}$, the model predicts $\sigma(t)\propto t^{-1/2}$. This mechanism is conditional on the stated scaling relationships and should be tested using direct measures of network activity and market liquidity. 

Historical Bitcoin price data produce volatility-exponent estimates that are broadly consistent with $\gamma$ near $1/2$ over several rolling-window lengths. The estimates are sensitive to temporal coarse-graining, however, and the use of overlapping windows prevents their native OLS standard errors from being interpreted as complete uncertainty estimates. The empirical results should therefore be regarded as a consistency check rather than a definitive test of a universal volatility exponent. 

Finally, we showed that an additional time-dependent multiplicative variance factor generally breaks exact Kelly invariance. Transaction-fee dispersion was considered as one illustrative scenario, rather than as a calibrated prediction of Bitcoin's future behaviour. More generally, the framework identifies a simple condition for age-invariant log-optimal allocation and provides a basis for testing whether that condition is approximately realised in non-stationary asset markets. Future work should test the proposed scaling relationships directly using network activity and market-liquidity data rather than price data alone.

\section*{Funding Statement}
This research did not receive any specific grant from funding agencies in the public, commercial, or not-for-profit sectors.

\section*{Declaration of Generative AI Assistance} Generative AI tools were used during manuscript development to assist with language editing, LaTeX formatting, code review, and the critical examination of mathematical and statistical arguments. The author independently reviewed, verified, and revised all resulting text, equations, code, interpretations, and references, and takes full responsibility for the content of the manuscript.

\bibliography{Econophysics}

@article{markowitz_portfolio_1952,
	title = {Portfolio {Selection}},
	volume = {7},
	issn = {0022-1082},
	url = {http://www.jstor.org/stable/2975974},
	doi = {10.2307/2975974},
	number = {1},
	urldate = {2021-01-16},
	journal = {The Journal of Finance},
	author = {Markowitz, Harry},
	year = {1952},
	pages = {77--91},
}

@article{santostasi_mechanistic_2026,
	title = {A mechanistic derivation of the {Bitcoin} price power law: {Network} adoption dynamics and generalised {Metcalfe} scaling},
	volume = {8},
	issn = {3050-5178},
	shorttitle = {A mechanistic derivation of the {Bitcoin} price power law},
	url = {https://www.sciencedirect.com/science/article/pii/S3050517826000675},
	doi = {10.1016/j.nls.2026.100172},
	urldate = {2026-08-26},
	journal = {Nonlinear Science},
	author = {Santostasi, Giovanni and Perrenod, Stephen},
	month = oct,
	year = {2026},
	pages = {100172},
}

@article{wheatley_are_2019,
	title = {Are {Bitcoin} bubbles predictable? {Combining} a generalized {Metcalfe}’s {Law} and the {Log}-{Periodic} {Power} {Law} {Singularity} model},
	volume = {6},
	issn = {2054-5703},
	shorttitle = {Are {Bitcoin} bubbles predictable?},
	url = {https://doi.org/10.1098/rsos.180538},
	doi = {10.1098/rsos.180538},
	number = {6},
	urldate = {2026-08-29},
	journal = {R. Soc. Open Sci.},
	author = {Wheatley, Spencer and Sornette, Didier and Huber, Tobias and Reppen, Max and Gantner, Robert N.},
	month = jun,
	year = {2019},
	pages = {180538},
}

@misc{noauthor_bitcoin_2026b,
	title = {Bitcoin {Price} {History}: {Download} {BTC} {Historical} {Data}},
	shorttitle = {Bitcoin {Price} {History}},
	url = {https://www.coingecko.com/en/coins/bitcoin/historical_data},
	urldate = {2026-09-06},
	journal = {CoinGecko},
	month = sep,
	year = {2026},
	howpublished = {\url{http://www.coingecko.com/en/coins/bitcoin/historical_data}},
	note = {Accessed: 2026-09-06},
}

@article{sornette_discrete-scale_1998,
	title = {Discrete-scale invariance and complex dimensions},
	volume = {297},
	issn = {0370-1573},
	url = {https://www.sciencedirect.com/science/article/pii/S0370157397000768},
	doi = {10.1016/S0370-1573(97)00076-8},
	number = {5},
	urldate = {2026-09-13},
	journal = {Physics Reports},
	author = {Sornette, Didier},
	month = apr,
	year = {1998},
	pages = {239--270},
}

@article{feigenbaum_discrete_1996,
	title = {Discrete scale invariance in stock markets before crashes},
	volume = {10},
	issn = {0217-9792},
	url = {https://www.worldscientific.com/doi/abs/10.1142/S021797929600204X},
	doi = {10.1142/S021797929600204X},
	number = {27},
	urldate = {2026-09-13},
	journal = {Int. J. Mod. Phys. B},
	publisher = {World Scientific Publishing Co.},
	author = {Feigenbaum, James A. and Freund, Peter G.o.},
	month = dec,
	year = {1996},
	pages = {3737--3745},
}

@book{bouchaud_theory_2003,
	title = {Theory of {Financial} {Risk} and {Derivative} {Pricing}: {From} {Statistical} {Physics} to {Risk} {Management}},
	isbn = {978-0-521-81916-9},
	shorttitle = {Theory of {Financial} {Risk} and {Derivative} {Pricing}},
	publisher = {Cambridge University Press},
	author = {Bouchaud, Jean-Philippe and Potters, Marc},
	month = dec,
	year = {2003},
}

@article{doyne_farmer_what_2004,
	title = {What really causes large price changes?},
	volume = {4},
	issn = {1469-7688},
	url = {https://doi.org/10.1080/14697680400008627},
	doi = {10.1080/14697680400008627},
	number = {4},
	urldate = {2026-09-13},
	journal = {Quantitative Finance},
	publisher = {Routledge},
	author = {Doyne Farmer, J. and Gillemot, László and Lillo, Fabrizio and Mike, Szabolcs and Sen, Anindya},
	month = aug,
	year = {2004},
	pages = {383--397},
}

@article{lillo_master_2003,
	title = {Master curve for price-impact function},
	volume = {421},
	copyright = {2003 Springer Nature Limited},
	issn = {1476-4687},
	url = {https://www.nature.com/articles/421129a},
	doi = {10.1038/421129a},
	number = {6919},
	urldate = {2026-09-13},
	journal = {Nature},
	publisher = {Nature Publishing Group},
	author = {Lillo, Fabrizio and Farmer, J. Doyne and Mantegna, Rosario N.},
	month = jan,
	year = {2003},
	pages = {129--130},
}

@article{merton_lifetime_1969,
	title = {Lifetime {Portfolio} {Selection} under {Uncertainty}: {The} {Continuous}-{Time} {Case}},
	volume = {51},
	issn = {0034-6535},
	shorttitle = {Lifetime {Portfolio} {Selection} under {Uncertainty}},
	url = {https://www.jstor.org/stable/1926560},
	doi = {10.2307/1926560},
	number = {3},
	urldate = {2026-09-13},
	journal = {The Review of Economics and Statistics},
	publisher = {The MIT Press},
	author = {Merton, Robert C.},
	year = {1969},
	pages = {247--257},
}

@book{maclean_kelly_2011,
	title = {The {Kelly} {Capital} {Growth} {Investment} {Criterion}: {Theory} and {Practice}},
	isbn = {978-981-4293-49-5},
	shorttitle = {The {Kelly} {Capital} {Growth} {Investment} {Criterion}},
	publisher = {World Scientific},
	author = {MacLean, Leonard C. and Thorp, Edward O. and Ziemba, W. T.},
	year = {2011},
}

@incollection{sornette_self-organized_2006,
	address = {Berlin, Heidelberg},
	title = {Self-{Organized} {Criticality}},
	isbn = {978-3-540-33182-7},
	url = {https://doi.org/10.1007/3-540-33182-4_15},
	doi = {10.1007/3-540-33182-4_15},
	urldate = {2026-09-13},
	booktitle = {Critical {Phenomena} in {Natural} {Sciences}: {Chaos}, {Fractals}, {Selforganization} and {Disorder}: {Concepts} and {Tools}},
	publisher = {Springer},
	editor = {Sornette, Didier},
	year = {2006},
	pages = {395--439},
}

@book{mantegna_introduction_1999,
	title = {Introduction to {Econophysics}: {Correlations} and {Complexity} in {Finance}},
	isbn = {978-1-139-43122-4},
	shorttitle = {Introduction to {Econophysics}},
	publisher = {Cambridge University Press},
	author = {Mantegna, Rosario N. and Stanley, H. Eugene},
	month = nov,
	year = {1999},
}

@book{bouchaud_trades_2018,
	title = {Trades, {Quotes} and {Prices}: {Financial} {Markets} {Under} the {Microscope}},
	isbn = {978-1-316-99888-5},
	shorttitle = {Trades, {Quotes} and {Prices}},
	publisher = {Cambridge University Press},
	author = {Bouchaud, Jean-Philippe and Bonart, Julius and Donier, Jonathan and Gould, Martin},
	month = mar,
	year = {2018},
}

@book{privman_finite_1990,
	title = {Finite {Size} {Scaling} and {Numerical} {Simulation} of {Statistical} {Systems}},
	isbn = {978-981-320-876-6},
	publisher = {World Scientific},
	author = {Privman, V.},
	month = jan,
	year = {1990},
}

@article{stanley_scaling_2001,
	title = {Scaling and universality in economics: empirical results and theoretical interpretation},
	volume = {1},
	issn = {1469-7688},
	shorttitle = {Scaling and universality in economics},
	url = {https://doi.org/10.1080/713666001},
	doi = {10.1080/713666001},
	number = {6},
	urldate = {2026-09-17},
	journal = {Quantitative Finance},
	publisher = {Routledge},
	author = {Stanley, H.E. and Plerou, V.},
	month = jun,
	year = {2001},
	pages = {563--567},
}

@article{clauset_power-law_2009,
	title = {Power-{Law} {Distributions} in {Empirical} {Data}},
	volume = {51},
	issn = {0036-1445},
	url = {https://epubs.siam.org/doi/abs/10.1137/070710111},
	doi = {10.1137/070710111},
	number = {4},
	urldate = {2026-09-17},
	journal = {SIAM Review},
	publisher = {Society for Industrial and Applied Mathematics},
	author = {Clauset, Aaron and Shalizi, Cosma Rohilla and Newman, M. E. J.},
	month = nov,
	year = {2009},
	pages = {661--703},
}

\end{document}